
\input amstex.tex
\documentstyle{amsppt}
\magnification\magstep1
\topmatter
\title SCATTERING MATRIX IN EXTERNAL FIELD PROBLEMS \endtitle
\author Edwin Langmann and Jouko Mickelsson \endauthor
\affil Department for Theoretical Physics, Royal Institute of
Technology, S-10044, Stockholm, Sweden \endaffil
\date September 7, 1995 \enddate
\endtopmatter

\NoBlackBoxes
\document

\redefine\a{\alpha}
\redefine\b{\beta}
\redefine\g{\gamma}
\redefine\G{\Gamma}
\define\gm{\bold g}
\redefine\e{\epsilon}
\define\Tr{\text{tr}}
\redefine\l{\lambda}

\define\R{\Bbb R}
\define\C{\Bbb C}

\define\um{\bold u}
\define\U{\text{U}}
\define\oprod{\prod}
\define\ee{\text{e}}
\define\cV{\widehat V}

\define\gA{g.A}
\define\gsA{g_s.A }

\define\cL{\Cal L}
\define\cA{\Cal A}
\define\cG{\Cal G}

\define\cO{\Cal O}


ABSTRACT: We discuss several aspects of second quantized scattering
operators $\hat S$ for fermions in external time dependent fields.  We
derive our results on a general, abstract level having in mind as a main
application potentials of the Yang--Mills type and in various dimensions.
We present a new and powerful method for proving existence of $\hat S$
which is also applicable to other situations like external gravitational
fields.  We also give two complementary derivations of the change of phase
of the scattering matrix under generalized gauge transformations which can
be used whenever our method of proving existence of $\hat S$ applies.  The
first is based on a causality argument i.e.\ $\hat S$ (including phase) is
determined from a time evolution, and the second exploits the geometry of
certain infinite-dimensional group extensions associated with the second
quantization of 1-particle operators.  As a special case we obtain a
Hamiltonian derivation of the the axial Fermion-Yang-Mills anomaly and the
Schwinger terms related to it via the descent equations, which is on the
same footing and traces them back to a common root.

\vskip 0.5in

1. INTRODUCTION

\vskip 0.3in
The main difficulty when quantizing fermions in higher than two
space-time dimensions in background (gauge) fields is that the
interaction term generically is too large to allow a naive
application of the standard methods of canonical quantization.
More precisely, if $\e$ is the sign of the 'free' Hamiltonian, then
only those one-particle operators $A$ are well-defined in the free Fock
space which  satisfy the condition that $[\e,A]$ is Hilbert-Schmidt.
For example, the minimal gauge interaction operator does not satisfy
this condition when the space-time dimension is higher than 2. The
same holds for gauge transformation operators which makes the
implementation of these operators somewhat tricky, [M2].

The one-particle time evolution operator can be constructed for example by
the Dyson expansion provided that the potential is smooth and appropriate
boundary conditions are satisfied.  However, the time evolution cannot be
quantized because of the remarks above.  The asymptotic scattering operator
$S$ is better behaving.  One can show that it satisfies the Hilbert-Schmidt
condition.  The existing proofs are rather involved, [P,R2].  In this paper
we shall give a much simpler proof using the methods introduced earlier for
the construction of the quantum gauge transformations and computation of
commutator anomalies, [M2].  The method is based on the observation that
the interaction Hamiltonians can be conjugated by unitary operators such
that the resulting equivalent Hamiltonians satisfy the Hilbert-Schmidt
condition with respect to fixed free Hamiltonian.  Moreover, we give an
effective method for an actual construction of such unitary conjugations,
as a function of (time dependent) background fields.  This method is very
general and does not use the specific properties of gauge interactions.  In
general, it applies to any bounded interactions such that its commutator
with the absolute value of the free Hamiltonian does not have worse
fall-off properties in the momentum space than the original operator.
Gravitational background fields can be also treated using a somewhat
modified form of the conjugation (Appendix A).

In sections 3 and 4 we discuss the determination of the phase of the
quantum scattering operator.  It is shown that the phase is uniquely
determined by causality (section 3), or, alternatively, by the geometric
structure of the central extension of the group of one-particle
(renormalized) time evolution operators (section 4). Our treatment relies
heavily on the theory of infinite-dimensional linear groups. Some of the
basic aspects of the theory of these groups in quantum field theory are
recalled on the way; for further reading we recommend [CR] and [GV].

\vskip 0.3in

2. EXISTENCE OF QUANTUM SCATTERING OPERATORS

\vskip 0.3in

Consider a family of Hamiltonians of the form $H_A(t)= D_0 + A(t)$ acting
in a one-particle Hilbert space $H$ where $t\mapsto A(t)$ is a smooth and
compactly supported ($t,t'\in\R$ here and in the following).  We assume
that $D_0$ is a self-adjoint operator and $D_0+A(t)$ is essentially
self-adjoint in the same domain for all $t$; the $A(t)$ are bounded
self-adjoint operators.  We study the time evolution equation $$i\partial_t
U_A(t,t')= H_A(t)U_A(t,t'),\quad U_A(t,t)= 1 .\tag2.1$$ Writing $V_A(t,t')=
\ee^{it D_0} U_A(t,t')\ee^{-it' D_0}$ we obtain an equivalent equation
$$i\partial_t V_A(t,t') = h_A(t) V_A(t,t'),\quad V_A(t,t)=1
\tag2.2$$ where $h_A(t)= \ee^{it D_0} A(t) \ee^{-it D_0}. $ Since $h_A(t)$
is bounded, this equation has a solution for all finite times given by
the Dyson expansion $$ V_A(t,t')= \sum_{n=0}^\infty V_n(t,t'),\quad
V_0(t,t')= 1,\quad V_{n+1}(t,t')= -i\int_{t'}^t ds\, h_A(s)V_n(s,t') \tag2.3$$
(it is easy to see that this series converges absolutely in the
operator norm, see Appendix B).

Let $\e= D_0/|D_0|.$ (This is well-defined even if zero is in the
spectrum of $D_0$ if we set $x/|x|=1$ and $-1$ for $x\geq 0$ and $x<0$,
respectively,  and use the spectral theorem of self-adjoint operators
[RS].)

The spectral decomposition $H= H_+\oplus H_-$ corresponding to the
splitting of the spectrum of $D_0$ to positive and negative part fixes
an irreducible representation of the canonical anticommutation
relations (CAR), uniquely defined up to unitary equivalence, in a Fock
space $\Cal F$ with a vacuum $|0>$ which is annihilated by the elements
$a^*(v_-)$ and $a(v_+),$ $v_\pm\in H_\pm$, of the CAR algebra
$$a^*(v) a(v') + a(v')a^*(v) = (v,v'),\tag2.4$$ and all the other
anticommutators are equal to zero. Let $\{e_n\}_{n\in\Bbb Z}$ be an
orthonormal basis in $H$ such that $\{e_n\}_{n\geq 0}$ span $H_+$ and
$\{e_n\}_{n<0}$ span $H_-$. Set $a_n= a(e_n)$ and $a^*_n = a^*(e_n).$
Fix the usual normal ordering for the products of creation and
annihilation operators by $:a^*_n a_m: = -a_m a^*_n$ if $n=m<0$ and all
the other products remain unchanged.

It is known that a bounded one-particle operator $X=(X_{nm})$ can be
canonically quantized as $$d\Gamma(X) = \sum X_{nm} : a^*_n a_m
:\tag2.5$$ iff $[\e, X]$ is Hilbert-Schmidt, [SS,A,CR].
This quantization is such that $[d\Gamma(X),a^*(v)]=a^*(Xv)$ for all
$v\in H$, and preserves the commutation relations of the Lie algebra
of linear operators on $H$ except for a complex valued cocycle
('Schwinger term'), see section 3. Similarly, a unitary operator $U$ on
$H$ can be second quantized to an operator $\Gamma(U)$ obeying
$\Gamma(U) a^*(v)\Gamma(U)^{-1}= a^*(Uv)$ if and only if $[\e, U]$ is
Hilbert-Schmidt [SS,R1].

If we have a time evolution with $[\e,A(t)]$ Hilbert Schmidt for all $t$,
then it is easy to see that $[\e,V_A(t,t')]$ is always Hilbert Schmidt (see
Appendix B), and this trivially implies that the scattering operator $$ S_A =
\lim_{t_f\to\infty}\lim_{t_i\to-\infty} V_A(t_f,t_i).\tag2.6 $$ can be
second quantized (note that due to our compactness assumption,
$S_A= V_A(T,-T)$ for some $T<\infty$).

In many interesting situations $[\e, A(t)]$ is not Hilbert-Schmidt, and
$[\e, V_A(t,t')]$ is not Hilbert-Schmidt either, and the canonical quantum
operator $\Gamma(V_A(t,t'))$ does therefore not exist.  Nevertheless the
scattering operator can be still second quantized in many such cases.  We
will proof below a general, abstract result for this.  As a motivation for
our abstract setting, we first discuss a special case.  We shall use some
basic facts about pseudodifferential operators (PSDO) [H]; see Appendix C
for notation.

We assume that spacetime is $M^n\times \R$ where $M^n$ is a
$n$-dimensional compact manifold with spin structure, and
$H= L^2(M^n)\otimes V$ where $V$ is a vector space carrying the spin
and color indices of the fermions. (Actually, the following discussion
applies also to noncompact situations like $M=\Bbb R^n$ but then one has to
assume suitable fall-off properties of the interaction as $|x|\to\infty.$
For example, in the case of a gauge interaction the requirement that the
vector potential and all its derivatives fall off faster than  $|x|^{-n/2}$
as $|x|\to\infty$ would be sufficient.)
Moreover, the free hamiltonian $D_0$
is a self-adjoint PSDO of order $\geq 1.$

We denote as $B_2$ the ideal of Hilbert Schmidt operators in the algebra of
bounded operators on $H.$ In case zero is in the spectrum of $D_0$ we
interpret $D_0^{-1}$ as $D_0(D_0^2+\lambda)^{-2}$ for some $\lambda>0$, and
similarly for $|D_0|^{-1}$.  We use this simplified notation since the
precise value of $\lambda$ is irrelevant (the essential regularizations
concern the ultraviolet and $\lambda$ is a harmless infrared regulator).
If not evident to the reader already at this point, this will become clear
in the following.

Denote as $O_{-k}$ the PSDOs of order $\leq -k.$ We assume that the free
hamiltonian $D_0$ satisfies $[|D_0|,a]\in O_{-k}$ for $a\in O_{-k}$ and for
each $k;$ this is the
case for example when $D_0$ is a Dirac operator.  The algebra of bounded
PSDO's is equal to $O_0.$ We state the basic properties of PSDO's, on a
compact manifold $M$ or in $\Bbb R^n$ with the asymptotic conditions
discussed above, which we shall need in the proof of the main theorem:

$$
\align
O_{-\ell}\subset O_{-k}
\quad \forall \ell > k &\qquad (i)\\
\forall a\in O_{-k}\text{ and } b\in O_0:\quad  ab,ba \in O_{-k}
&\qquad (ii)\\
\forall a\in O_{-k}:\quad [|D_0|, a] \text{ and } |D_0|^{\pm 1} a|D_0|^{\mp
1}\in O_{-k}  &\qquad (iii)  \tag 2.7\\
\forall a\in O_{-k}:\quad |D_0|^{-1}a
\text{ and } a |D_0|^{-1} \in O_{-k-1}  &\qquad (iv)\\
\exists p<\infty: O_{-p}\subset B_2 . &\qquad (v)
\endalign
$$

In fact, we shall prove a result implying that $S_A$ can be quantized
whenever there are operator families $\{ O_{-k} \}_{k= 0,1,\ldots}$ with
the properties (2.7) and $A\in O_0$.  The idea is to construct a time
dependent family of operators $T(A)=T_t(A)$ for a regularization at the one
particle level [M2] i.e.\ consider the modified time evolution
$T_t(A) U_A(t,t')T_{t'}(A)^{-1}$ which can be second quantized even if
$U_A(t,t')$ cannot.  It is easy to see that the latter is generated by the
Hamiltonian $H'_A(t)=
T_t(A) H_A(t) T_t(A)^{-1} + i (\partial_tT_t(A)) T_t(A)^{-1} )=
D_0+A'(t)$ where $$ A'(t) = i(\partial_t T_t(A))T_t(A)^{-1} -
[D_0,T_t(A)]T_t(A)^{-1} + T_t(A) A(t) T_t(A)^{-1} .  \tag2.8$$ Our strategy
thus is to choose $T(A)$ in such a way that $A'$ is better behaved than the
original interaction $A$, i.e.\ that $[\e,A']\in B_2$.  Note that a
conjugation $T_t(A)$ which becomes the identity as $|t|\to \infty$ does not
alter the scattering matrix, $S_{A'} = S_A.$

All differentiations of operators in the following are meant with
respect to the operator norm.

\proclaim{Definition 2.9} Let $D_0$ be a self-adjoint operator such that
there are operator families $\{ O_{-k} \}_{k= 0,1,\ldots}$ with the
properties (2.7).  We call an interaction $A$ {\bf regular} (w.r.t.  to
$D_0$ and $\{ O_{-k} \}_{k= 0,1,\ldots}$) if $A(t)$ and the derivatives
$(\partial_t)^k [\e,A(t)]$, $k= 1\ldots p$, are in $O_0$ for all $t\in
\R$. We denote the set of all such interactions as $\cA$.
\endproclaim

\proclaim{Theorem 2.10} For all interactions $A(t)\in\cA$, there is a
family of unitary operators $T_t(A)$ differentiable in $t$ and such that
the transformed time evolution $V_{A'}(t,t')$, $A'$ eq.\ (2.8),
can be second quantized, $[\e,V_{A'}(t,t')]\in B_2$ for all $t,t'\in\R$.
Moreover, $T(A)$ can be chosen local in time, i.e.\ $T_t(A)=1$ if $A(t)=0$ and
$(\partial_t)^k A(t)=0$ for $k= 1\ldots,p$.\endproclaim

\proclaim{Corollary 2.11} For all interactions $A(t)\in\cA$ compactly
supported in $t$, the scattering operator $S_A$ exists and can be second
quantized, $[\e,S_A]\in B_2$.
\endproclaim

Before proving this theorem, we give one other example of such operator
families $\{ O_{-k} \}_{k= 0,1,\ldots}$ with the properties (2.7).  It is
easy to see that it generalizes the PSDO setting for the Dirac operator
$D_0$ on a compact spin manifold $M^n$ discussed above: Let $D_0$ be such
that for some $\lambda\in\R$, $(D_0+\lambda)^{-1}$ exists and is in the
Schatten class $B_p=\{a\in B|\, (a^*a)^{p/2} \text{ trace class}\}$ for
some $p<\infty$.  Let $O_0$ be all bounded operators $a$ such that
$(D_0+\lambda)^{-\ell}a(D_0+\lambda)^{\ell}$ is bounded for all integers
$\ell$.  One can then check that the operator families $O_{-k}=\left\{a\in
O_0|\, a |D_0|^k \text{ is bounded}\right\}$, $k=0,1,\ldots$, satisfy
(2.7).

\demo{Proof of Theorem}
We say a map $a:\R\to O_{-k}$ is $C^r$ if it is $r$ times differentiable
with all derivatives $\partial_t^\ell a $, $\ell =
1,2,\ldots r$, continuous maps $\R\to O_{-k}$. We first prove the following
key lemma providing the recipe for constructing $T(A)$:

\proclaim{Lemma 2.12} Let $A:\R\to O_0$ such that
$[\e,A]: \R \to O_{-k}$ be $C^r$ with $r\geq 1$.  Then $A'(t),$ defined by
(2.8), with
the unitary operator $$ T_t(A)= \ee^{\alpha(t)} ,\quad \alpha(t)=
-\frac18\left(|D_0|^{-1}[\e,A(t)]+[\e,A(t)]|D_0|^{-1}\right) \tag 2.13 $$
defines a map
$A':\R\to O_0$ such that $[\e,A']$ maps $\R$ into $O_{-k-1}$ and is
$C^{r-1}$.
\endproclaim

\demo{Proof of Lemma}  We write $A'(t) =  A'_1(t) +
A'_2(t) $ where
$$A'_1 =  A + [D_0,\alpha] $$
is the leading terms
in an expansion in powers of $|D_0|$, and
$$A'_2 =
-i T(A)^{-1}\partial_t(T(A)-1)  + T(A)^{-1}[D_0,T(A)-\alpha-1] +
T(A)^{-1}[A,T(A)-1] $$
is the rest. In the following we refer to maps
$a: \R\to O_{-k}$ also as $a\in O_{-k}$ etc.

Since obviously $\alpha$ and $\partial_t(\alpha)$ are in $O_{-k-1}$,
$A'_2\in O_{-k-1}$ trivially follows from $\alpha D_0, D_0\alpha \in O_{-k}$
and $$ T(A)-1= \alpha T_1= T_1\alpha ,\quad T(A)-1-\alpha = \alpha^2 T_2 =
T_2 \alpha^2 $$ where $ T_{1,2}$ and $\partial_t(T_1)$ all are in $O_0$.

The nontrivial part thus is to show that  $[\e,A'_1]\in O_{-k-1}$. This
can be seen by the following calculation,
$$
\align
[\e,A_1'] =  [\e,A]  -\frac{1}{8}\left[\e,\left[D_0,|D_0|^{-1}[\e,A(t)] +
[\e,A]|D_0|^{-1}\right] \right] \\ =
\frac18\left(8[\e,A] -\left[\e, \e[\e,A] - |D_0|^{-1}[\e,A(t)]|D_0|\e
+ \e |D_0|[\e,A(t)]|D_0|^{-1} - [\e,A(t)]\e \right] \right) \\=
\frac18\left(8[\e,A] - 4[\e,A] - 2|D_0|[\e,A]|D_0|^{-1} -
2|D_0|^{-1}[\e,A]|D_0| \right) \\=
\frac14\left(|D_0|^{-1}\left[|D_0|,[\e,A]\right] - \left[|D_0|,[\e,A]\right]
|D_0|^{-1} \right)
\endalign
$$
where we used $[\e,A] =  -\e [\e,A]\e$ and
$\e D_0^{\pm 1} =  D_0^{\pm 1}\e =  |D_0^{\pm 1}|$.
Thus
$$[\e,A_1']  =
\frac{1}{4}\left[|D_0|^{-1},\left[|D_0|,[\e,A]\right]\right]
$$ which is in $O_{-k-1}$ by
definition. If we replace $D_0^{-1}$ by $D_0(D_0^2+\lambda)^{-1}$
a similar calculation leads to the same conclusion,
$$
[\e,A_1'] =  \frac14\left[
|D_0|(D_0^2+\lambda)^{-1},\left[|D_0|,[\e,A]\right] \right] +
\frac12\lambda\left((D_0^2+\lambda)^{-1}[\e,A]+[\e,A](D_0^2+\l)^{-1}\right).
$$
This proves our Lemma.
\enddemo

We can apply this method successively: Starting from some interaction $A_0=
A$ such that $[\e,A]\in O_0$ we get a new interaction $A_1=A'$ using the
conjugation $T(A),$ with $[\e,A_1]\in O_{-1}.$ We can then insert $A_1$ as
an argument to $T(\cdot)$ and obtain an unitary operator $T(A_1).$ This
defines again a new interaction $A_2=A_1'$ such that $[\e,A_2]\in O_{-2}.$
Continuing this way we obtain, after $p$ steps, an unitary operator
$T^{(p)}(A) = T(A_{p-1})\dots T(A_0)$ such that the time evolution for the
operator $T_t^{(p)}(A) U(t,t')T_{t'}^{(p)}(A)^{-1}$ is determined by an
interaction $A_p$ such that $[\e,A_p(t)]\in O_{-p}$ for all $t$.  For
sufficiently big $p$ the new interaction satisfies the Hilbert-Schmidt
condition, and thus the corresponding scattering operator can be second
quantized.  Since $T_t^{(p)}(A)$ by construction is equal to the identity
for times $t$ where $A(t)$ and all its $t$-derivatives vanish,
the latter scattering operator is equal to $S_A$.  This implies theorem 2.10.

\enddemo

\demo{Remark 1} As a particular case, our result gives the existence of the
scattering operators for Dirac (or Weyl) fermions in external Yang-Mills
fields, on a compact space manifold $M$ or on $\Bbb R^n$ with sufficient
fall-off
properties for the vector potential as $|x|\to\infty$. Here our
discussion above implies $p>n/2$, but for $n$ odd one can show that
actually $p=(n-1)/2$ is already sufficient (e.g.\ for $n=1$ no
regularization is necessary).
\enddemo

\demo{Remark 2} We stress the Hilbert-Schmidt property of the
scattering operator since only this is of primary interest for quantum
field theory.  However, our the argument above shows that usually
$[\e,S_A]$ is much better behaved: e.g.\ in the fermion-Yang-Mills case it
is in \it all \rm Schatten classes $B_q$
for $q>0$.  (This follows from $O_{-k}\subset B_{\frac{n+1}{k}}$.)
\enddemo

\vskip 0.3in

3. PHASE OF QUANTUM SCATTERING OPERATOR: CAUSAL APPROACH

\vskip 0.3in
In the previous section we have shown that the one-particle scattering operator
$S$ satisfies the Hilbert-Schmidt condition for $[\e,S]$ and therefore it can
be promoted to an unitary operator $\hat S=\Gamma(S)$ in the Fock space
$\Cal F.$
However, by this the operator $\hat S$ is uniquely defined only up to a phase.
In this section we show that the regularization for the time evolution
operators in the previous section fixes the phase in a natural causal manner.

We denote the group of unitary operators $U$ on $H$ with $[\e,U]$
Hilbert-Schmidt as $U_1$.  All $U\in U_1$ can be second quantized, and the
second quantization $\Gamma(U)=\Gamma(U^{-1})^{-1}$ of $U\in U_1$ is
unique up to a phase (= element in $\U(1)$) which implies that for some
local (near the unit element in $U_1$, e.g.) choice of of phases $$
\Gamma(U)\Gamma(V) = \chi(U,V) \Gamma(UV)\quad \forall U,V\in U_1, \tag 3.1
$$ where $\chi:U_1\times U_1\to \U(1)$ is only defined locally
(a derivation of an explicit, locally valid formula of $\chi$ is given in
[PS]; in the second quantization setting see also [L1]).
The latter is a nontrivial local 2-cocycle providing a central extension
$\hat U_1$ of $U_1$ by $\U(1).$ Similarly the (complexification of the) Lie
algebra $\um_1$ of $U_1$ contains all bounded operators $X$ on $H$ with
$[\e,X]$ Hilbert-Schmidt, and its second quantization $\um_1\ni X\to
d\Gamma(X)=d\Gamma(X^*)^*$ gives a representation of a central extension
$\hat\um_1=\um_1\oplus \C$ of $\um_1$, $$
[d\Gamma(X),d\Gamma(Y)]=d\Gamma([X,Y]) + c_L(X,Y), \tag 3.2 $$ with a Lie
algebra 2-cocycle [Lu] $$ c_L(X,Y)=\frac14 \Tr \e [\e,X][\e,Y] \tag 3.3 $$
which is the infinitesimal version of the Lie group 2-cocycle $\chi$
above.  It is possible to choose phases such that
$$\Gamma(\ee^{i X}) = \ee^{i d\Gamma(X)}\quad \forall X=X^*\text{ close to
$0\in \um_1$}
\tag 3.4
$$ (the existence of the $\ee^{i d\Gamma(X)}$ as a unitary operator follows
from Stone's theorem [RS] since $d\Gamma(X)$ is self-adjoint [CR]).  This
equation actually is true for all $X\in\um_1$, but it fixes the phase of
$\Gamma(U)$ for only for $U\in U_1$ sufficiently close to the identity
where local bijectivity of the exponential mapping is guaranteed.  We will
assume this phase convention in the following.  Then (3.2) implies $$
\Gamma(\ee^{-i\delta s X}\ee^{-i\delta t Y}\ee^{i\delta s X}) =
\ee^{\delta s\delta t \, c_L(X,Y)}\,
\Gamma(\ee^{-i\delta s X})\,\Gamma(\ee^{-i\delta t Y})\, \Gamma(\ee^{i\delta
s X})+ \cO(\delta s^2,\delta t^2) \tag 3.5 $$ for all $X,Y\in\um_1$ and
sufficiently small $\delta s,\delta t\in\R$ (to see this, use (3.4) and
expand both sides of this equation in powers of $\delta s$ and $\delta t$).

We now consider a time evolution $V_A(t,t')=V(t,t')$ defined in eq.\ (2.2)
with $h_A=h:\R\to \um_1$ smooth and compactly supported.  We first consider
the simple case where $h(t)\in\um_1$ so that $V(t,t')\in U_1$ for all
$t,t'\in\R$.  As shown in the last section, many interesting cases can be
brought to this simplest situation using the conjugation by a family of
operators $T(A)$ (we will discuss this in more detail further below).

We first note the essential group property of the time evolution,
$$ V(t,t')V(t',t'')= V(t,t'')\quad \forall t,t',t''\in\R,
\tag 3.6
$$ which follows from (2.2); it is this what we mean by \it causality\rm.
Somewhat parallel to our discussion, the use of the causality condition
in the renormalization of a quantum field theory has been stressed by
Scharf and his coworkers, [S].

To construct the second quantization of the scattering operator $S=S_A$
(2.6) including the phase, we first second quantize the time evolution.
The naive guess $\Gamma(V(t,t'))$ for this is not right since this is not a
time evolution: it does not obey an equations similar to (3.6) due to the
Schwinger term $\chi$ in (3.1) which gives nontrivial contributions in
general.  One can, however, define $\cV(t,t') =
\lim_{N\to\infty}\cV^{(N)}(t,t')$ with $$
\cV^{(N)}(t,t')=\oprod_{N\geq \nu\geq 1}
\Gamma\left(V(t_\nu,t_{\nu-1})\right),\quad t_\nu= t'+\frac{(t-t') \nu}{N}
\tag 3.7
$$ where $\oprod_{N\geq \nu\geq 1}F(t_\nu,t_{\nu-1})$ is the ordered
product $F(t_N,t_{N-1})F(t_{N-1},t_{N-2})\cdots F(t_2,t_1)$ for any
operator valued function $F$ on $\R\times\R$.  This is a time evolution by
construction, and with (3.1) $$\cV(t,t')= \eta(t,t')\Gamma(V(t,t')) \tag
3.8a$$ where $\eta$ a phase valued function on $\R\times\R$ which can be
explicitly computed in terms of $\chi$ (for $V(t,t')$ in some
neighborhood of the identity) [L1].  This allows to calculate the
scattering operator $\hat S=\cV(T,-T)$ including phase as follows (here and
in the following we assume that $T$ is big enough so that $h(t)$ vanishes
for $|t| > T/2$, say): choose some partition $t_0=-T<t_1<\cdots < t_n=T$
of the time interval $[-T,T]$ such that all $V(t_{i},t_{i-1})$ are in the
neighborhood of the identity for which $\eta(t,t')$ is defined.  Then $$
\hat S = \oprod_{n\geq i\geq 1} \eta(t_{i},t_{i-1}) \Gamma(V(t_{i},t_{i-1}))
\tag 3.8b
$$
can be shown to be independent of which particular partition is chosen.

\demo{Remark 1} We note our formulas (3.8a,b) still do not fix the phase of
$\hat S$ completely since the function $\eta(t,t')$ is unique only up to
$$
\eta(t,t') \mapsto \exp\left(-i \int_{t'}^t d\bar t\, E(\bar t)
\right)\eta(t,t') \tag 3.9
$$ with $E$ a smooth real-valued functions on $\R$.  This is due to the
ambiguity of the second quantization map $\um_1\ni X\to d\Gamma(X)$ which can
be changed by smooth, linear functions $b:\um_1\mapsto \C$ with
$b(X^*)=b(X)^*$ and $b(0)=0$.  A shift $d\Gamma(X)\to d\Gamma(X)+b(X)$
changes (3.3) by a trivial 2-cocycle, $c_L(X,Y) \to c_L(X,Y) - b([X,Y])$,
and this implies (3.9) with $E(t)= b(h(t))$.
\enddemo

\demo{Remark 2}
Since $h(t)\in\um_1$ for all $t$, the second quantized Hamiltonian $\hat
h(t) = d\Gamma(h(t))$ (in the interaction picture) always exists, and it
should be the generator of the second quantized time evolution $\cV(t,t')$.
Moreover, the ambiguity (3.9) of the phase of $\cV(t,t')$ corresponds to a
shift $\hat h(t) \to \hat h(t) +E(t)$ which physically amounts to a change
of the zero-point energy.  It would be difficult to construct $\cV(t,t')$
directly from $\hat h(t)$ since the latter is unbounded which makes the
existence of a Dyson series nontrivial.  This technical problem is avoided
in our approach above.
\enddemo

In the following we are interested in the change of the second quantized
time evolution operator under transformations $$ V(t,t')\mapsto (g.V)(t,t')
\equiv g(t)V(t,t')g(t')^{-1} \tag 3.10 $$ where $g:\R\mapsto U_1$ where
$g(t)$ is assumed to be sufficiently smooth and
such that $g(t)=1$ for $|t| > T/2$.  We will derive an explicit formula
for the gauge anomaly of the time evolution, $$\lambda(t,g)\equiv
\Gamma(g(t))^{-1}\, \widehat{(g.V)}(t,-T)\,
\cV(-T,t), \tag 3.11
$$ which is a phase factor according to our discussion above (since the
r.h.s.  is the second quantization of $g(t)^{-1}(g.V)(t,-T)V(-T,t)$ equal to
the identity).  Especially, $\lambda(g)\equiv
\lambda(T,g)$ is the change of the quantum scattering operator $\hat S$
under the transformation $g$.

We first consider only infinitesimal gauge transformations
$g(t)=\ee^{-i \delta s X(t)}$ for $\delta s\to 0$. We calculate $\lambda(t,g)$
as $\lim_{N\to\infty} \lambda^{(N)}$ where
$$
\lambda^{(N)} =
\Gamma(g(t))^{-1}
\left\{\oprod_{N\geq \nu\geq 1}
\Gamma\left(g(t_\nu)V(t_\nu,t_{\nu-1}) g(t_{\nu-1})^{-1} \right) \right\}
\left\{ \oprod_{1\leq \nu\leq N}
\Gamma\left(V(t_{\nu-1},t_\nu)\right)\right\}
$$
with $t_\nu = -T+ (t+T) \nu/{N}$.  Now (3.1) implies $$
\Gamma\left(g(t+\delta t)V(t+\delta t,t)g(t)^{-1}\right) =
\lambda^{(t+\delta t,t)}(g)
\Gamma(g(t+\delta t))\Gamma(V(t+\delta t,t))\Gamma(g(t))^{-1}
$$
for some phase factors $\l^{(t+\delta t,t)}$, and we explicitly see that the
various factors $\Gamma(g(t_\nu))$ and $\Gamma(g(t_{\nu-1}))^{-1}$ cancel each
other leaving only phase factors. Using $V(t+\delta t,t) \simeq
\ee^{-i\delta t h(t)}$,  $g(t+\delta t)\simeq
\ee^{-i\delta s X(t)}$ and (3.5), we get $\lambda^{(t+\delta
t,t)}(g)\simeq \ee^{\delta s\delta t\, c_L(X(t),h(t))}$
(`$\simeq$' means `equal up to irrelevant higher order terms in
$\delta s$ and $\delta t$'). Thus $\lambda^{(N)}$ is just the exponent of a
Riemann sum, and in the limit $N\to\infty$
$$
\lambda\left(t,\ee^{-i\delta s X}\right)
= \exp\left(\delta s\int_{-T}^t d\bar{t}\, c_L(X(\bar{t}),h(\bar{t}))
\right) + \cO(\delta s^2) .
$$

We now consider the case of finite gauge transformations $g(t)$ and
introduce a homotopy $g_s(t)$, $0\leq s\leq 1$,  smoothly deforming it
to the identity,
$$
g_1(t)=g(t)\text{ and }  g_0(t)=1\quad \forall t,
\quad g_s(t)=1\quad\text{ for } |t| > T/2.
\tag 3.12
$$ To be specific, we first restrict ourselves to gauge transformations
$g(t)=\ee^{-iX(t)}$ with $X(t)\in\um_1$ for all $t$, and $g_s(t)=g(t)=\ee^{-i
s X(t)}$.  We define $V_s(t,t')\equiv (g_s.V)(t,t')$ and $$\lambda_{s,s'}
\equiv \Gamma(g_s(t))^{-1}
\cV_s(t,-T)\cV_{s'}(-T,t)\Gamma(g_{s'}(t))$$
so that $\lambda(g)=\lambda_{1,0}$. We observe
that these phases have the group property,
$\lambda_{s,s'}\lambda_{s',s''}= \lambda_{s,s''}$ for
all $0\leq s,s',s''\leq 1$, thus we can evaluate
$\lambda(g)$ as $\lim_{M\to\infty} \lambda_M$ where
$$
\lambda_M= \oprod_{M\geq \mu \geq 1}\lambda_{s_{\mu},s_{\mu-1}},\quad
s_\mu = \frac{\mu}{M}.  $$ Now $\lambda_{s+\delta s,s}$ is the change of
phase of $\cV_{s}(T,-T)$ under an infinitesimal gauge transformation
$g_{s+\delta s}(t)g_{s}(t)^{-1}\simeq \ee^{-i \delta s X_s(t)}$ and thus equal
to $\exp(\delta s \int_{-T}^t d\bar{t}c_L(X_s(\bar{t}),h_s(\bar{t})))$ with
$$
h_s(t)=i \left\{\partial_tV_s(t)\right\}  V_s(t)^{-1},
\quad
X_s(t)=i \left\{\partial_s V_s(t)\right\} V_s(t)^{-1},
\quad V_s(t)=g_s(t)V(t,-T)\tag 3.13
$$ (we used $g_s(-T)=1$).  Again $\lambda_M$ becomes the
exponential of a Riemann sum, and in the limit $M\to\infty$ we obtain
\proclaim{Theorem 3.14}
$$
\lambda(t,g)=
\exp\left(\int_0^1 ds \int_{-T}^t d\bar{t} \,
c_L\left(X_s(\bar{t}),h_s(\bar{t})\right)\right) .$$ \endproclaim

Note that this result was derived for the special homotopy $g_s(t)=\ee^{-i
sX(t)}$, but our derivation can be immediately generalized to arbitrary
gauge transformations $g(t)$ and homotopies $g_s(t)$ (sufficiently smooth
in $s$ and $t$) obeying (3.12).  For $t\geq T$, $\lambda(t,g)=\lambda(g)$
(3.14) is then actually independent of the homotopy chosen (this follows
from its definition (3.11) which does not depend on the homotopy).  For
intermediate times $-T<t<T$ this is not true.  The reason is that
then the phase of the implementors $\Gamma(g(t))$ in (3.11)
depends on the homotopy: our derivation
above implies that this phase has to be chosen such that
$$
\Gamma(g(t)) = \lim_{M\to\infty} \oprod_{M\geq \mu\geq 1}
\Gamma(g_{s_\mu}(t)g_{s_{\mu-1}}(t)^{-1})\, ,\quad s_\mu=\frac{\mu}{M},
$$
and this coincides with our phase convention (3.4) only for homotopies
$s\mapsto g_s(t)=\ee^{-isX(t)}$.

\demo{Remark 3} Our derivation of (3.14) above was given for 1-parameter
groups in $U_1$ for simplicity, but the result immediately generalizes to
$GL_1$ which is the group of all (not only unitary) invertible operators
$U$ on $H$ with $[\e,U]$ Hilbert Schmidt: eq.  (3.14) remains true for
$h(t)\in\um_1$ not self-adjoint and $g(t)\in GL_1$.  The technical problem
for proving this more general result by the method above is that
$\ee^{-i\delta t d\Gamma(X)}$ is unbounded if $d\Gamma(X)$ is not
self-adjoint, thus one has to be careful with the domains of operators (the
latter could, however, be handled by methods described in [GL]).  Our
alternative derivation of (3.14) in the next section is for $GL_1$ and
bypasses such domain questions.
\enddemo

We consider now time evolutions generated by Hamiltonians $H_A(t)=D_0+A(t)$
with $A(t)\in\cA$ (cf.\ definition 2.9) and generalized gauge
transformations $$A(t)\to \gA(t) = i (\partial_t g(t)) g(t)^{-1} -
[D_0,g(t)]g(t)^{-1} + g(t)A(t) g(t)^{-1} \tag 3.15$$ so that
$U_{\gA}(t,t')= g(t)U_A(t,t')g(t')^{-1}.$ We denote the group of all $g(t)$
which leave $\cA$ invariant as $\cG$.  Note that $\cG$ contains all $g(t)$
sufficiently smooth in $t$ (i.e.\ $C^{p+1}$), which are unitary operators
in $O_0$ for all $t$.  We also introduce the Lie algebra $Lie\cG$ of $\cG$.
In the following, all $A$ are in $\cA$, all $g,g',g''$ in $\cG$, and all
$X,Y,Z\in Lie\cG$, except when stated otherwise.  As before, we assume all
these functions are trivial for $|t|>T/2$.

By theorem 2.10, there exist appropriate
regularization operators $T(A)$ and $T(\gA)$ such that $A'$ and $(\gA)',$
defined in (2.8), all lead to time evolutions which can be second quantized
i.e.\ they are always in $\um_1$.  This also implies that the operators
$T_t(\gA) g(t) T_t(A)^{-1} = U_{\gA}(t,-T)U_A(-T,t)$ all are in $U_1$, and
thus $$
\Gamma_t(g;A) \equiv \Gamma\left(g'_A(t) \right)\, ,\quad
g'_A(t)\equiv \ee^{itD_0}\,T_t(\gA)\, g(t) \,T_t(A)^{-1}\,\ee^{-itD_0}
\tag 3.16
$$ always exist.  These unitary operators have the natural interpretation
as implementors of the generalized gauge transformations $g$ at fixed time
$t$.  They are local in time i.e.\ only depend on $g$, $A$ and
$t$-derivatives thereof, at time $t$. We observe that they obey the
relations $$
\Gamma(g';{\gA})\, \Gamma(g;A) = \chi(g',g;A)\,\Gamma(g'g;A). \tag 3.17a
$$
were we have dropped the common time argument $t,$ with
$$
\chi_t(g',g;A)= \chi\left((g')'_{\gA}(t),g'_A(t)\right)
\tag 3.17b$$
defined locally (this follows from (3.1)).
Note that (3.17a) and associativity of the operator product
imply the 2-cocycle relation
$$
\chi(g'', g'g; A) \,\chi(g',g ;A) =  \chi(g'',g' ;\gA) \,\chi(g''g',g ; A).
\tag 3.18
$$

Our construction above can now be used to calculate
$$
\lambda(t,g;A)\equiv \Gamma_t(g;A)^{-1}\,
\widehat{V_{(\gA)'}}(t,-T)\, \widehat{V_{A'}}(-T,t) \tag 3.19
$$ which we define as the change of the quantum time evolution
$\widehat{V_{A'}}$ under the generalized gauge transformation $g$.
We immediately get the formula
$$\lambda(t,g;A) =
\exp{\left(\int_0^1 ds\int_{-T}^t d\bar{t}\, c_L\left( \left\{\partial_s
V_s(\bar{t})\right\} V_s(\bar{t})^{-1},
\left\{\partial_t V_s(\bar{t})\right\} V_s(\bar{t})^{-1}\right)
\right)} \tag 3.20a$$
with $$ V_s(t)= \ee^{itD_0}\, T_t(\gsA) \, g_s(t) \,U_A(t,-T)\, \ee^{iTD_0}
\tag 3.20b$$ and $g_s(t)\in\cG$ a homotopy interpolating between
$1$ and $g(t)$.  Similarly as discussed above after theorem 3.14, for
$t\geq T$
(but in general not for intermediate times $-T<t<T$) this formula is
independent of the homotopy $s\mapsto g_s(t)$ chosen.

We observe that these phases are connected with the
Schwinger terms in (3.17b) via the relation
$$
\lambda(t,g';\gA)\, \lambda(t,g;A)\,\chi_t(g',g;A) = \lambda(t,g'g;A)
\tag 3.21
$$
(this follows from a simple calculations using the definition (3.19) and
$(g'g).A=g'.(g.A)$,
$$
\align
\Gamma_t(g';\gA)\,\lambda(t,g';\gA)\,  \Gamma_t(g;A)\,\lambda(t,g;A)
\qquad \qquad \qquad \qquad \qquad \\
= \widehat{V_{(g'g.A)'}}(t,-T)\,\widehat{V_{(\gA)'}}(-T,t)\,
\widehat{V_{(\gA)'}}(t,-T)\,\widehat{V_{A'}}(-T,t)
\qquad \qquad \qquad \qquad \qquad \\
= \Gamma_t(g'g;A)\,\lambda(t,g'g;A),
\qquad \qquad \qquad \qquad \qquad
\endalign
$$ and inserting (3.17a)).  According to our derivation, this equation is
valid only locally (i.e.\ $g(t)$ and $g'(t)$ close to identity).

Especially for $t=T$, $\lambda(g;A)\equiv \lambda(T,g;A)$ is equal to the
change of the quantum scattering matrix $\hat S_A$ under the transformation
$g$, and equation (3.21) reduces to the 1-cocycle relation,
$\lambda(g';\gA)\, \lambda(g;A) = \lambda(g'g;A)$ (since $\chi(1,1;0)=1$).
The physical meaning of $\lambda(g;A)$ is as follows.  We recall that the
$\log$ of the vacuum expectation value of $\hat S_A$ is equal to the
Minkowskian action of the fermions in the time dependent external field
$A$, thus $\log \lambda(g;A)$ is the change of the latter by the
generalized time dependent gauge transformation $g(t)$.  Especially for
infinitesimal transformations $g(t)=1-i\delta s X(t) +\cdots$ it gives the
generalized gauge anomaly $Anom(X;A)= \left.\frac{d}{d
s}\log\lambda(\ee^{-isX};A)\right|_{s=0}$.  We obtain
$$Anom(X;A) =
\int_{-T}^Tdt\,\bar\omega^1(X(t),A(t)),\quad
\bar\omega^1(X(t),A(t)) = c_L\left(X_A'(t), h_{A'}(t)\right)
\tag 3.22$$
where  $h_{A'}(t)=\ee^{itD_0}A'(t)\ee^{-itD_0}$ with $A'$ given in
eq.\ (2.8),  and
$$ X'_A(t)\equiv \ee^{itD_0}
\left(\left\{\cL_X T_t(A)\right\} T_t(A)^{-1} + T_t(A) X(t) T_t(A)^{-1}
\right)\ee^{-itD_0}
\tag 3.23$$
is in $\um_1$ for all $A\in\cA$ and $X\in Lie\cG$; we introduced the Lie
derivative acting on functionals $f$ on $\cA$ as $$\cL_Xf(A) =
\left.i\frac{d}{d s}f(\ee^{-isX}A)\right|_{s=0}.$$  Similarly, the
infinitesimal version of (3.17a,b) is [M2]
$$ [G(X;A),G(Y;A)] = G([X,Y];A)+ S(X,Y;A) \tag 3.24a $$
where $G(X;A) = \cL_X + d\Gamma(X'_A)$ are
implementors of infinitesimal gauge transformations and
$$ S_t(X,Y;A) = c_L(X'_A(t),Y'_A(t)) \tag 3.24b $$
a Schwinger term satisfying the 2-cocycle
relation $\cL_X S(Y,Z;A) + S(X,[Y,Z];A)+cycl.=0$ (the latter is the
infinitesimal version of (3.18) and also follows from the Jacobi identity).

Especially, if we consider the Yang-Mills case and infinitesimal chiral
gauge transformations, $Anom(X;A)$ is just the axial gauge anomaly and
$S(X,Y;A)$ the Schwinger term appearing in the commutators of the chiral
Gauss' law generators $G(X;A)$. We thus have obtained a Hamiltonian
derivation of these two different manifestations of the gauge anomaly in
a Hamiltonian framework which traces them back to a common root, i.e.\
the 2-cocycle $c_L$ in (3.2).

It is interesting to consider also the infinitesimal version of the
equation (3.21) which can be written as
$$
\delta \bar \omega^1 +\partial_t S = 0 \tag3.25
$$
where
$$(\delta \omega^1)(X,Y;A) = \cL_X \omega^1(Y;A) - \cL_Y
\omega^1(X;A) - \omega^1([X,Y];A)$$
is defined on functions $\omega^1$ on $Lie\cG\times\cA$.  To interpret this
equation, we recall that the above mentioned fermion-Yang-Mills anomalies
are connected by descent equations [Z]: the axial anomaly on a $n+1$
(even) dimensional space-time manifold $M^{n+1}$ is the integral of a
$(n+1)$- (de Rham) form $\omega^1_{n+1}(X;A)$ over $M^{n+1}$; it depends on
one infinitesimal gauge transformations $X$ and the Yang-Mills field A.  The
corresponding Schwinger term is on $n$ dimensional space $M^n$ and an
integral of a $n$-form $\omega_n^2(X,Y;A)$ over $M^n$
depending on two infinitesimal gauge transformations $X,Y$ and $A$.
Embedding $M^n$ in $M^{n+1}$, the descent equations are
$\delta\omega^1_{n+1} +
d\omega^2_n=0$ where $\delta$ is defined as above and
$n$ is the usual exterior differentiation of de Rham forms.  Setting
$M^{n+1}=M^{n}\times\R$ and $\bar\omega^1=\int_{M^n}\omega^1_{n+1}$ and
$S=\int_{M^n}\omega^2_n$, one exactly obtains our equation (3.25).  We
thus have obtained an explicit field theory derivation of this descent
equation for all odd dimensions $n$ in the Hamiltonian framework.  We
stress, however, that our eq.  (3.25) is not restricted to the Yang-Mills
case but in fact is much more general.

\demo{Remark 4} As just mentioned, fermion-Yang-Mills anomalies are
local de Rham forms, whereas our formulas (3.22) for the axial anomaly and
(3.24b) for the Schwinger term are not explicitly local in space.  In the
Yang-Mills case one can prove, however, that they cohomologous to local de
Rham forms.  General arguments and mathematical techniques for showing this
by explicit calculations have been given recently, [M2,LM,M4].
Nevertheless it would be interesting to explicitly do this latter
calculations for all dimensions.  In this paper we will only sketch the
simplest case $n=1$ (end of next section).
\enddemo

In the next Section we will give a different, more geometric approach to
the phase of the scattering operator where the path independence of the
anomaly becomes evident.  Another important benefit in the geometric
approach is that we can easily compute the cohomology class of the anomaly
without going to the details of the renormalization $T(A)$.

\vskip 0.3in
4. THE QUANTUM PHASE AND PARALLEL TRANSPORT

\redefine\gg{\text{{\tt g}}}

\vskip 0.3in

Let $\hat G$ be a central extension of a Lie group $G$ by
$\Bbb C^{\times}.$ The Lie algebra $\hat\gm$ of $\hat G$ is a vector
space direct sum $\gm\oplus\Bbb C.$ Let $\pi$ be the projection on
the second summand and let $\theta=\gg^{-1}d\gg$ be the left
Maurer-Cartan one-form. We can then define a complex valued one-form $\phi$ on
$\hat G$ by $\phi=\pi(\theta).$ This is a connection form in the
principal $\Bbb C^{\times}$ bundle $\hat G\to G.$ Its curvature is a
left invariant two-form on $G$ given by $\omega(X,Y)=c(X,Y),$ where
left invariant vector fields $X,Y$ on $G$ are identified as elements
of the Lie algebra and $c$ is the 2-cocycle on $\gm$ defining the
central extension,
$$ [(X,\l),(Y,\mu)]=([X,Y],c(X,Y)).\tag4.1$$

Recall that $GL_1$ is the group of invertible linear transformations
$\gg:H\to H$ such that $[\e,\gg]$ is Hilbert-Schmidt and $U_1$ its unitary
subgroup.  Let us apply the above remarks to $G=U_1,$ and to the Lie
algebra cocycle $c_L$ (3.3) arising when promoting the one-particle
operators to operators (2.5) in the fermionic Fock space, as discussed in
the last section.

The central extension $\widehat{GL}_1$ is a nontrivial $\Bbb C^{\times}$
bundle over the base $GL_1,$ [PS].
The elements of the group $\widehat{GL}_1$ (containing the unitary
subgroup $\hat U_1$) can be thought of equivalence classes of pairs
$(\gg,q),$  where $\gg\in GL_1$ and $q:H_+\to H_+$ is an invertible
operator such that $a-q$ is a trace-class operator,
$$\gg=\left(\matrix a&b\\c&d\endmatrix\right).\tag4.2$$
We have assumed that ind$\,a=0.$ If this is not the case, the subspace
$H_+$ must be either enlarged or made smaller by a suitable
finite-dimensional subspace in order to achieve ind$\,a=0.$
The equivalence relation is determined by $(\gg,q)\sim (\gg',q')$ if
$\gg=\gg'$ and det$(q'q^{-1})=1.$ Thus the fiber of the extension is $\Bbb C^
{\times}$ and it is parameterized by (the nonexisting ) determinant of $q.$

The product is defined simply $(\gg,q)(\gg',q')=(\gg\gg',qq').$ Near the unit
element in $G$ we can define a local section $\gg\mapsto (\gg,a),$ [PS].
Denoting
$$\gg^{-1}=\left(\matrix \a&\b \\ \g & \delta\endmatrix\right)$$
we can write the connection form as
$$\phi_{\gg,q}=\Tr [(\gg^{-1}d\gg)_a-q^{-1}dq]
=\Tr [\a da +\b dc -q^{-1}dq].\tag4.3$$
The curvature of this connection is
$$\omega= \Tr (d\b d\g )\tag4.4$$
and is easily checked to agree with $c_L$ in (3.3).

We compute the parallel transport determined by the connection in the
range of the local section. Let $\gg(t)$ be a path in $GL_1,$
$-T\leq t\leq T,$ with $\gg(-T)=1.$ The lift $(\gg(t),q(t))$ is parallel if
$$0= \phi_{\gg(t),q(t)}(d\gg,dq)= \Tr [\a(t)
a'(t)+\b(t)c'(t)-q(t)^{-1}q'(t)].$$
Thus the parallel transport, relative to the trivialization $\gg\mapsto
(\gg,a),$ along the path $\gg(t)$ in the base is
accompanied with the multiplication by the complex number
$$\exp\{-\int_{-T}^T\Tr [(\a(t)-a(t)^{-1})a'(t)+\b(t)c'(t)]dt \}\tag4.5$$
in the fiber $\Bbb C.$

Formally,
$$\Tr\, q^{-1}q'=\Tr[\a a'+\b c']$$
and so
$$ \text{det} \, q(T) = \exp\int_{-T}^T \Tr[\a(t)a'(t)+\b(t)c'(t)]dt$$
and also
$$ \text{det}\,a(T)= \exp\int_{-T}^T \Tr \,a(t)^{-1}a'(t)dt.$$
Individually, the traces in these two expressions do not converge,
but putted together the trace converges and gives
$$\text{det}(a(T)q(T)^{-1})=\exp\{ \int_{-T}^T
\Tr [(\a -a^{-1})a'+\b c']dt\}.\tag4.6$$
Note that the exponent diverges outside of the domain of the local section,
reflecting the fact that det$\,a(T)=0$ outside of the domain.

We can now apply the above results to the 'renormalized' one-particle
time evolution operators $\gg(t)= V_{A'}(t) =
\ee^{itD_0} T_t(A)U_A(t,-T) \ee^{iTD_0}.$
For all times $t$, these are elements
of the group $U_1.$ On the other hand, in the Fock representation of
$\widehat{GL}_1$ these correspond to elements $\widehat{V_{A'}}(t)$ in the
central
extension $\hat U_1.$ The phase of the quantum time evolution operator
is then uniquely given by the parallel transport described above.

The Minkowskian effective action is by definition the vacuum expectation
value of the quantum scattering operator  $\hat S_A.$ The vacuum is invariant
under the free time evolution $\exp(itD_0)$ and taking into account the
assumption that the interaction has essentially compact support in time,
we can write
$$Z(A) = <0| (V_{A'}(T),q(T))|0> . \tag4.7$$
The vacuum expectation value is given by a simple formula, [PS], [M3],
$$<0| (\gg,q)|0>=\text{det}(aq^{-1})\tag4.8$$
and therefore the parallel transport (4.5) (with respect to the given local
trivialization) is equal the effective action $Z(A).$

The above formalism can be applied for computing the gauge anomaly in the
space-time formalism starting from the commutator anomaly (3.3).  Let
$g(t)\in\cG$ be a time-dependent gauge transformation such that at $t=\pm T$ it
is equal to the identity.  The change in the phase of the effective action
is now $$\l(g;A)=\exp\left(\int_{\g} \phi\right)$$ where $\g$ is the closed
loop in $U_1$ obtained by first following backwards in time from $T$ to
$-T$ the time evolution $U_{A'}(t)$, following then the gauge transformed
time evolution operators $g(t)U_{A'}(t)$ back from $-T$ to $T$.  The
parallel transport around a closed loop can be written as an integral of
the curvature $\omega$ over a surface $S$ enclosed by the loop $\g.$
By construction, the gauge anomaly $\l$ satisfies the 1-cocycle
condition $\l(gg';A)=\l(g;g'.A)\l(g';A).$

Joining $g(t)$ to the identity by a homotopy $g_s(t),$
$0\leq s\leq 1,$ and writing $V_s(t)=g_s(t)U_{A'}(t)$ we get
$$\log\lambda(g;A)= \int_S c_L(\partial_t V V^{-1},\partial_s VV^{-1})=
\frac14 \int \Tr\, \e [\e,\partial_t V V^{-1}][\e,\partial_s
V V^{-1}].\tag4.9$$
This result agrees with (3.14).
For infinitesimal gauge transformations $g_s(t)\simeq 1 -is
X(t)+\dots$ we get axial anomaly (3.22) as discussed in the last Section.

Let us complete the calculation for 1+1 spacetime dimensions in the case of
chiral fermions in external Yang-Mills field.  Now the chiral Hamiltonian
on the circle $S^1$ acting on one-component spinors is $H(t)=-i\frac
{\partial }{\partial
x}-A_+,$ where $A_+=A_0+A_1.$ We now use that for $n=1$ one can choose
$T_t(A)=1$ independent of $A$ (see our remark 1 at the end of section 2).
Thus, applying (3.22) derived either from (3.20) or (4.9), we get
$$Anom(X;A)=
\frac14 \int_{-T}^T dt\, \Tr\, \e [\e,A_+][\e, X(t)]
=\frac{1}{2\pi i}\int_{-T}^T dt
\int_{S^1}dx\, \Tr\, A_+(t,x) \frac{\partial}{\partial x} X(t,x).$$
Here we have used the general formula
$$\frac14 \Tr\,\e[\e,X][\e,Y]=
\frac{1}{2\pi i} \int_{S^1} dx\, \Tr\, X \partial_x Y\tag4.10$$
valid for smooth
multiplication operators $X,Y$ on the unit circle.  Up to a coboundary (= a
gauge variation of the local functional $\propto \int \Tr A_+ A_1 $)
this form of the anomaly is equal to the standard form of the
two-dimensional chiral anomaly $$Anom_s(X;A)=
\frac{1}{4\pi i}\int_{S^1\times\R} \Tr \,A dX.\tag4.11$$

We finally note that this same equation also allows to calculate the
Schwinger term (3.24b), $$ S(X,Y;A) = \frac{1}{2\pi i}
\int_{S^1}\Tr\, X d Y \tag 4.12
$$ which actually is independent of $A$.  This is the Kac-Moody cocycle and
also the Schwinger term related to the axial anomaly (4.11) via the descent
equations, as discussed at the end of the last section.

\vskip 0.3in

\it The cohomology class of the anomaly in dimensions $n+1>2$ \rm

\vskip 0.3in
The group $GL_p$ consists of all bounded invertible operators (4.2)
in $H=H_+\oplus H_-$ such that the off-diagonal blocks $b,c$ are in the
Schatten ideal $B_{2p}.$
For any $p\geq 1 $ the group $GL_p$ contracts to the subgroup $GL_1,$ [Pa].
On the other hand, in $GL_1$ one can produce cohomologically equivalent
cocycles
$c_p \sim c_L$ such that $c_p$ extends from $GL_1$ to $GL_p.$ These are
relevant for understanding the gauge group action in space-time dimension
$n+1>2.$ The static gauge transformations are elements of $GL_p$ for $p>
n/2.$ For example, when $n=3$ the gauge group $\Cal G_n=Map(M^n,G)\subset GL_2$
and one has, [MR],
$$c_2(X,Y;f) = \frac18 \Tr\,[\e,f]f^{-1}[[\e,X],[\e,Y]],\tag4.13$$
where $c_2(X,Y;f)$ is the value of a 2-form on $GL_2$ at a point $f$ to the
directions of the left invariant vector fields (= Lie algebra elements) $X,Y.$
This formula has been generalized for arbitrary $p$, [FT,L4].

In order to fix the cohomology class of the one-cocycle $\l(\gg;A)$ it is
sufficient to look how $\l$ winds around the circle when a family $f(t,s)$
of time dependent gauge transformations wraps around a closed surface $S$
(parameterized by $s,t$) in the group $\Cal G_n$ of static gauge
transformations.
This follows from the fact that the cohomology class of any two-form is
determined by giving its integral over all closed two-cycles.
The winding number is given by the integral of the curvature $c_L$ around the
surface $S$ in $GL_1$ defined by the family of gauge transformed renormalized
evolution operators.

For any fixed potential $A$ and a homotopy $f(t,s)$ of time dependent gauge
transformations we have a map $S\to GL_p$ given by $(t,s)\mapsto f(t,s) U(t),$
where $U(t)$ is the nonrenormalized time evolution operator determined by $A.$
The renormalization $T(A)$ does not change the homology class of the surface
$S$ in $GL_1\subset GL_p$ since $T$ is defined over a contractible
parameter space. It follows that the integral over a \it closed \rm
surface $S$ of the curvature on $GL_1$ is given by the integral of
$c_p$ of the nonrenormalized operators $f(t,s)U(t).$ Furthermore,
the surface $(t,s)\mapsto f(t,s)U(t)$ contracts to $(t,s)\mapsto f(t,s).$
This follows from the fact that each component of $U_p$ is simply connected
and so $(t,s)\mapsto U(t)$ is contractible.
Therefore, the final result for the anomaly around a closed surface is
$$ \int_{f(t,s)} c_p.$$
In the case $M=S^1$ ($p=1$) this gives
$$\align \frac14 \int_S \Tr\,\e [\e,(\partial_t f) f^{-1}]& [\e,(\partial_s f)
f^{-1}] =\frac{1}{2\pi i} \int_{S\times S^1} \Tr(\partial_t f)f^{-1}\partial_x(
(\partial_s f)f^{-1}) \\
&= \frac{i}{12\pi}   \int_{S\times S^1} \Tr (f^{-1}df)^3,\endalign$$
where we have used (4.10) in the first step and performed integration by
parts in the second step.
In dimension $n=3$ we insert $X=i(\partial_t f)f^{-1}$ and $Y=i(\partial_s f)
f^{-1}$ into (4.13). A similar calculation gives, using a three dimensional
equivalent of (4.10), [L2],
$$ \frac{1}{240 \pi^2}   \int_{S\times {M^3} } \Tr (f^{-1} df)^5$$
and so on in higher dimensions.
This agrees with the integral of the
two form over $S$ in a gauge orbit obtained by the descent equations.
For example, in three space dimensions the form is
$$ \frac{-i}{24\pi^2}\int_{M^3} \Tr dA(XdY-YdX)\tag4.14$$
which gives the commutator anomaly in three space dimensions, [M1, FS].

\vskip 0.3in
\bf Conclusions: \rm  In section 2 we gave a new proof for the existence
of the second quantized fermionic scattering operator in external Yang-Mills
fields. The proof is valid also in a more abstract setting of generalized
gauge interactions in the spirit of Connes' noncommutative geometry.
In section 3 we derived a formula for the phase of the scattering operator
and its gauge variation from the concept of causality by using the local
2-cocycle on the group $GL_1.$ In section 4 we gave an alternative geometric
derivation using a connection on the global group extension $\widehat{GL}_1.$
A constructive interpretation for the descent equations was given in the
hamiltonian framework, linking the anomaly of the Minkowskian effective
action to the Schwinger terms. This is complementary to the standard
approach which starts from the euclidean functional determinant.

\vskip 0.3in
ACKNOWLEDGEMENT
\vskip 0.3in

We wish to thank S.N.M. Ruijsenaars for discussions and critical
remarks. We also want to thank G. Marmo and P. Michor for inviting us to
the Erwin Sch\"odinger Institute where part of this work was
done.

\vskip 0.3in
APPENDIX: THE CASE OF EXTERNAL METRIC

\vskip 0.3in
Let $g=(g_{ij}(x,t))$ be a time-dependent metric tensor in $\Bbb R^n$.
We assume that space and time has been foliated by a choice of the time
coordinate such that the space coordinates $x_1,\dots,x_n$ are orthogonal
with respect to the time $t,$ i.e.\ $g_{0i}=g_{i0}=0$ for $1\leq i\leq n.$
The Dirac equation is written as
$$i g_{00} \partial_t \psi= \g^k h_{kj} (i\partial_j +\G_j)\psi
\simeq g_{00} D_h \psi     \tag A1$$
where $h_{kj}(x)$ are the components of an oriented orthonormal basis
in $(\Bbb R^n,g),$
$$h_{kj} h_{mj} = g_{km}.\tag A2$$
The matrices $\G_j$ are the components of the spin connection (defined by the
Levi-Civita connection of $g$), taking values in the Lie algebra of the
spin group $Spin(n).$

We assume that the deviation of the metric $g$ from the euclidean metric has
only compact support in space and time. Furthermore, $g(x,t)$ is assumed to be
smooth. If the dimension $n=2N+1$ then the $g$'s are $2^N\times 2^N$ complex
matrices with the property
$$\g^i\g^j +\g^j \g^i = 2\delta_{ij}.\tag A3$$
The Lie algebra of $Spin(n)$ is spanned by the commutators $[\g_i,\g_j].$
If $n=3$ the $\g$-matrices are just the $2\times 2$ Pauli matrices which are
also
the generators of the Spin group $Spin(3)=SU(2).$

The principal symbol of the Dirac Hamiltonian is $\g^k h_{kj} p_j.$ The
complete symbol is the sum of the principal symbol and of a symbol of order
zero in the momenta.

Because for any given pair $q,p$ of nonzero vectors there is a rotation $R$
such that $q=Rp,$ there exists an element $B(p,x)\in Spin(n)$ such that
$$B \g^k p_k B^* = \l(x,p) \g^k h_kj p_j .\tag A4$$
Here the product is a matrix product, no momentum space differentiation is
involved, and the scale factor $\l$ is the ratio of the euclidean lengths of
the vectors $p=\g^k p_k$ and $q=\g^k h_{kj}p_j.$ Both $\l$ and $B$ are
homogeneous functions of order zero in momenta.

At the first sight it appears that it is not possible to construct $B$ as a
continuous function of the $h$ field, the apparent obstruction being the
hairy ball theorem: For a given direction $q$ one can always choose a
rotation $R_q$ such that $R_q\cdot p= q,$ but $R_q$ is not a continuous
function of $q$ when $n$ is odd and at least equal to 3. However, here we
can profit from the information encoded in the matrix $h.$

The set of all orthogonal transformations which takes
$p$ to $q=h\cdot p$ (up to a scale) form a fiber $P_{p,q}$ in a principal
bundle $P$
with base $X=GL_+(n,\Bbb R)\times S^{n-1},$ consisting of the pairs
$(h,p/|p|)$,
and the fiber is isomorphic with $SO(n-1);$ the '+' refers to matrices with
positive determinant. The base contracts to
$X'=SO(n)\times
S^{n-1}$ (by the Cartan decomposition). On the other hand, over $X'$ the bundle
$P$ is trivial, the trivialization being given by $(h,p)\mapsto h.$ Thus
$P$ is trivial. We choose a trivialization $(h,p)\mapsto R(h,p).$
We choose $B(h,p)\in Spin(n)$ which projects down to $R\in SO(n).$ There is a
$\Bbb Z_2$ ambiguity in the choice which does not bother us since the
transformation law for the Dirac operator is quadratic in $B.$

If we compute the left-hand side in (A4) with the complete star product instead
of the matrix product, we generate symbols of order less than or equal to zero.
Thus we have proven the following lemma:
\proclaim{Lemma} There is a function $B(h)$ of the basis $h$ taking values
in the group of invertible PSDO's of order zero such that $B^*\g^k h_{kj} p_j
B$  differs
from $\l(x,p)^{-1} \g^k  p_k$ by an operator of order zero. \endproclaim

The unitarily equivalent Hamiltonian $B'=B^* D_h B$ has then the property that
$[\epsilon, B^* D_h B]$ is a PSDO of order zero.  One can now apply the
recursive method in section 2 to obtain the  renormalization operator
$T=T(A)$ where now $A=B'-D_0$ and $D_0= \g^k p_k.$ This method applies as
well to the case of
combined background gauge and gravitational interactions.

\vskip 0.3in

APPENDIX B: ESTIMATES ON DYSON SERIES

\vskip 0.3in
We consider here the Dyson series (2.3) solving the time
evolution equation (2.2) with $h_A(t)= \ee^{it D_0} A(t) \ee^{-it D_0}.$
If $A(t)$ is bounded for all $t$, one can easily prove by induction that
$$
||V_n(t,t')|| \leq \frac{1}{n!}\left(\int_{t'}^t dr ||A(r)|| \right)^n
$$
which shows that (2.3) converges in the operator norm $||\cdot ||$
for all $t,t'\in\R$.

Similarly, if $[\e,A(t)]$ is Hilbert Schmidt for all
$t$, then the Hilbert Schmidt norm of $V_n$ can be estimated as
$$
|| [\e,V_n(t,t')] ||_2\leq
\int_{t'}^t dr ||[\e,A(r)]||_2
\frac{1}{(n-1)!}\left(\int_{t'}^t dr ||A(r)|| \right)^{n-1}
$$
showing that $[\e,V(t,t')]$ is also Hilbert Schmidt.

\vskip 0.3in

APPENDIX C: PSEUDODIFFERTIAL OPERATORS (PSDO)

\vskip 0.3in
To fix our notation we summarize here the basic definitions
and facts about PSDOs [H].
A PSDO $A$ on the Hilbert space $L^2(M^n)\otimes V$, $M^n$ a smooth
manifold  and $V$ a finite dimensional vector
space, is given locally by its symbol $a(x,p)= \sigma(A)(x,p)$ which is a
smooth matrix- (gl$(V,V)$-) valued function of the local coordinates
$x\in U\subset \Bbb R^{n}$  and momenta $p\in\Bbb R^{n},$ [H].
The action of $A$ on a section $\psi$ with support in $U$ is given as
$$(A\psi)(x)= \frac{1}{(2\pi)^{n/2}} \int  a(x,p) \hat\psi(p)\ee^{-ip\cdot x}dp
\tag C1$$
where $\hat\psi$ is the Fourier transform of the function $\psi:U\to V,$
$$\hat\psi(p) = \frac{1}{(2\pi)^{n/2}} \int \ee^{ix\cdot p} \psi(x) dx.$$

We shall consider \it the restricted class \rm of PSDO's which admit an
asymptotic expansion of the symbol as
$$a(x,p) \sim a_{k}(x,p)+a_{k-1}(x,p) +a_{k-2}(x,p)+\dots$$
where $k$ is an integer and each $a_j$ is a homogeneous matrix valued
function of the momenta, of order $j,$ with $|a_j|\sim |p|^j$ as
$\sqrt{p_1^2 +\dots + p_n^2}=  |p|\to\infty.$ The order of such a PSDO is
ord$a= k$.

The asymptotic expansion for the product of two PSDO's
is given by the formula
$$a*b \sim \sum \frac{(-i)^{|m|}}{m!} \left[(\partial_p)^m a(x,p)\right]
\left[ (\partial_x)^m
b(x,p)\right],\tag C2$$
where the sum is over all sets of nonnegative integers $m= (m_1,\dots,m_{n}=
),$
$|m|= m_1+\dots +m_{n},$ ${\partial_x}^m= (\frac{\partial}{\partial x_1})
^{m_1}\dots(\frac{\partial}{\partial x_n})^{m_{n}},$ etc., and
$m!= m_1!\dots m_{n}!$.

The order of $a*b$ is  equal to the sum of ord$a+$ ord$b$ since
the leading term in $a*b$ is just the matrix product $ab$ of the symbols.

The symbol of a massless Dirac operator $D_A$ in an external vector potential
$A$ is $\g^k (p_k + A_k)$ where $\g_i \g_j +\g_j \g_i = g_{ij}$ are the
Dirac gamma matrices and $g=(g_{ij})$ is the metric tensor. The symbol
for the square $D_A^2$ is $p^2 + $ lower order terms in $p$ and therefore
the symbol of $|D_A|$ is $|p|+$ lower order terms. From this follows, using
(A2), that the symbol of $[|D_A|, B]$ is $\frac{p_k}{|p|}\frac{\partial}{
\partial x_k} b(x,p)+$ terms of order ord$\,B$ for any PSDO $B$ with symbol
$b.$ In particular,
the order of $[|D_A|,B]$  is at most equal to the order of $B.$

On a compact manifold of dimension $n$ a PSDO is trace class if its order
is strictly less than $-n$ and it is Hilbert-Schmidt if the order is
$< -n/2.$ In $\Bbb R^n$ one has to assume in addition that the symbol is either
compactly supported in $x$ or at least the asymptotic behavior of the
symbol and its derivatives at
$|x|\to \infty$ is as $|x|^{-k}$, where $k> n$ in case of trace class
operators and $k > n/2$ for Hilbert-Schmidt operators. In $\Bbb R^n$ the
trace (when it exists) of a PSDO is simply given as
$$\Tr \, A = \frac{1}{(2\pi)^n} \int \Tr\, a(x,p) dpdx.$$

\vskip 0.3in
REFERENCES

\vskip 0.3in

[A] H. Araki: Bogoliubov automorphims and Fock representations of canonical
anticommutation relations. In: \it Contemporary Mathematics\rm , vol. 62,
American Mathematical Society, Providence (1987)

[CR] A. L. Carey  and S.N.M. Ruijsenaars:
On fermion gauge groups, current algebras and Kac-Moody algebras.
Acta Appl. Mat. \bf 10 \rm, 1 (1987)

[FS] L.  Faddeev and S.  Shatasvili: Algebraic and Hamiltonian methods in
the theory of non-Abelian anomalies.  Theoret.  Math.  Phys.  \bf 60 \rm,
770 (1984)

[FT] K. Fujii and M. Tanaka: Universal Schwinger cocycle
and current algebras in $(D+1)$- dimensions: geometry and physics.
Comm. Math. Phys. \bf 129 \rm, 267 (1990)

[GL] H. Grosse and E. Langmann: A super-version of quasi-free second
quantization. I. Charged particles.  J. Math. Phys. \bf 33 \rm, 1032
(1992)

[GV] J.M. Gracia-Bondia and J.C. Varilly: QED from the spin representation.
J. Math. Phys. \bf 35 \rm, 3340 (1994)

[H] L. H\"ormander: \it The Analysis of Linear Partial
Differential Operators III. \rm Springer-Verlag, Berlin (1985)

[L1] E. Langmann: Cocycles for boson and fermion Bogoliubov
transformations. J. Math. Phys. \bf 35 \rm, 96 (1994)

[L2] E.  Langmann: Fermion current algebras and Schwinger terms in
(3+1)--Dimensions.  Comm.  Math.  Phys.  \bf 162 \rm, 1 (1994)

[L3] E. Langmann: Noncommutative integration calculus. J. Math. Phys. \bf
36 \rm, 3822 (1995)

[L4] E. Langmann: Descent equations for Yang--Mills anomalies in
noncommutative geometry. \tt hep-th/9508003 \rm

[LM] E. Langmann and J. Mickelsson:
(3+1)-dimensional Schwinger terms and non-commutative geometry.
Phys. Lett. B \bf 338 \rm, 241 (1994)

[Lu] L.-E. Lundberg: Quasi-free 'second quantization'. Commun. Math. Phys.
\bf 50 \rm, 103 (1976)

[M1] J. Mickelsson: Chiral anomalies in even and odd dimensions. Commun.
Math. Phys. \bf 97 \rm, 361 (1995)

[M2] J. Mickelsson:  Regularization of current algebra.
In: \it Contraint Systems and Quantization. \rm Ed. by
Colomo, Lusanna, and Marmo. World Scientific, Singapore (1994);
Wodzicki residue and anomalies of current algebras.
In: \it Integrable Systems and Strings. \rm Ed. by
A. Alekseev et al. Springer Lecture Notes in Physics (1994);
Hilbert space cocycles as representations of (3+1)-D current algebras.
Lett. Math. Phys. \bf 28 \rm, 97 (1993)

[M3] J. Mickelsson: \it Current Groups and Algebras. \rm Plenum Press,
London and New York (1989)

[M4] J. Mickelsson: Schwinger terms, gerbes, and operator residues.
\tt hep-th/9509002 \rm

[P] J. Palmer: Scattering automorphisms of the Dirac field.
J. Math. Anal. Appl. \bf 64 \rm, 189 (1978)

[Pa] R.S. Palais: On the homotopy type of certain groups of operators.
Topology \bf 3 \rm, 271 (1965)

[PS] A. Pressley, G. Segal: \it Loop Groups. \rm Clarendon Press, Oxford (1986)

[R1] S.N.M. Ruijsenaars: On Bogoliubov transformations
for systems of relativistic charged particles.
J. Math. Phys. \bf 18 \rm, 517 (1977)

[R2] S.N.M.  Ruijsenaars: Charged particles in external fields.  I.
Classical theory.  J.  Math.  Phys.  \bf 18 \rm, 720 (1976).  Gauge
invariance and implementability of the $S$-operator for spin-$0$ and
spin-$\frac{1}{2}$ particles in time-dependent external fields.  J.  Funct.
Anal.  \bf 33 \rm, 47 (1979)

[RS] R. Reed and B. Simon:
\it Methods of Modern Mathematical Physics I. Functional
Analysis. \rm  Academic Press, New York (1968)

[S] G. Scharf: \it Finite Quantum Electrodynamics. \rm Springer Verlag, Berlin
(1989)

[SS] D. Shale and W.F. Stinespring: J. Math. and Mech. \bf 14 \rm, 315
(1965)

[Z] B. Zumino: Cohomology of gauge groups: cocycles and
Schwinger terms. Nucl. Phys. B \bf 253 \rm, 477 (1985)

\vskip 0.3in

\enddocument